\documentclass[prd,preprintnumbers,amsmath,amssymb,nofootinbib]  {revtex4}
\usepackage[dvips,final]{graphicx}
\usepackage{amsmath}
\usepackage{amssymb}
\usepackage{url}
\usepackage{subfigure}
\usepackage{textcomp}
\usepackage{graphicx}
\usepackage{bm}
\usepackage{dcolumn}
\usepackage[usenames]{color}

\usepackage{epstopdf}

\usepackage{graphicx}
\newcommand{\beq}{\begin{equation}}
\newcommand{\eeq}{\end{equation}}
\newcommand{\bea}{\begin{eqnarray}}
\newcommand{\eea}{\end{eqnarray}}
\newcommand{\bi}{\begin{itemize}}
\newcommand{\ei}{\end{itemize}}

\def\beq{\begin{equation}}
\def\eeq{\end{equation}}
\def\bea{\begin{eqnarray}}
\def\eea{\end{eqnarray}}

\makeatletter
\newcommand*{\rom}[1]{\expandafter\@slowromancap\romannumeral #1@}
\makeatother

\begin{document}

\title{ A Non-minimally Coupled Potential for Inflation and Dark Energy after Planck 2015: A Comprehensive Study }

\author{Mehdi Eshaghi$^{1,2}$}
\email{eshaghi249@gmail.com}

\author{Moslem Zarei$^{3,5}$}
\email{m.zarei@cc.iut.ac.ir}

\author{Nematollah Riazi$^{4}$}
\email{n_riazi@sbu.ac.ir}

\author{ Ahmad Kiasatpour $^{1}$}
\email{akiasat@sci.ui.ac.ir}

\affiliation{ $^1$ Department of Physics , Faculty of Science, University of Isfahan, Isfahan, 81746-73441, Iran}

\affiliation{ $^2$ Astrophysics Sector, SISSA, Via Bonomea 265, I-34136 Trieste, Italy}

\affiliation{ $^3$ Department of Physics, Isfahan University of Technology, Isfahan 84156-83111, Iran}

\affiliation{$^4$ Department of Physics, Shahid Beheshti University, Tehran 19839, Iran}

\affiliation{$^5$ School of Astronomy, Institute for Research in Fundamental Sciences (IPM),P.~O.~Box 19395-5531,Tehran, Iran}


\date{\today}

 \begin{abstract}

In this work we introduce a new plateau-like inflationary model including a quadratic scalar potential coupled non-minimally to gravity. This potential has a dominant constant energy density at early times which can realize successful inflation. It also includes an infinitesimal non-zero term $V_0$ responsible for explaining dark energy which causing the universe to expand accelerating at the late time. We show that this model predicts small tensor-to-scalar ratio of the order of $r\approx 0.01$ which is fully consistent with Planck constraints. Using the lower and upper bounds on reheating temperature, we provide additional constraints on the non-minimal coupling parameter $\xi$ of the model. We also study the preheating stage predicted by this kind of potentials using numerical calculations.
 \end{abstract}

\maketitle

\section{Introduction}

In physical cosmology, cosmic inflation is the extremely rapid exponential expansion of the early universe, driven by a negative-pressure vacuum energy density
 \cite{Guth,Linde1,Linde2}. Inflation is an elegant way to resolve the horizon, flatness, and monopole problems in standard cosmology. Usually a scalar field with a nearly flat potential, called inflaton, is introduced to produce such accelerating exponential expansion. In this manner, inflaton field slowly rolls down its potential and then automatically produces an inflation phase. The inflationary stage is controlled by the slow roll parameters
 \begin{equation}
  \epsilon = \frac{1}{2\kappa^{2}}\left(\frac{V'}{V}\right)^{2} \:\:\:\:\:
  \textrm{and}\:\:\:\:\:\eta= \frac{1}{\kappa^{2}}\frac{V''}{V}~,
 \end{equation}
 where $\kappa=M_{P}^{-1}=(8\pi G)^{1/2}$, $V\equiv V(\phi)$ is the inflaton potential and the prime denotes derivative with respect to
$\phi$. During inflation the slow roll parameters are small $\epsilon,\:\eta\ll 1$ and inflation ends by the breakdown of these slow roll conditions. For single field inflationary models one can calculate the spectral index $n_s$ and the tensor-to-scalar ratio  $r$ in terms of slow roll parameters as $n_{s} =  1-6\epsilon+2\eta$ and $r =16\epsilon$, respectively. The recent Planck data estimate the scalar spectral index to be $n_{s}=0.968\pm 0.006$ and establish an upper bound on the tensor-to-scalar ratio at $r < 0.09$ \cite{Planck:2015xua}. By using Planck+BKP data, quadratic inflationary model and natural inflation are ruled out but models including a simple linear potential and fractional-power monomials and also models that have sub-Planckian field evolution are however still allowed by the data \cite{Planck:2015xua,Ade:2015lrj}. In general, single field inflationary models with plateau-like shapes are more favored by Planck+WMAP data \cite{Ijjas:2013vea}. However, there are still a large number of possible models satisfying the Planck+WMAP constraints \cite{Martin:2014vha}.

It is clear that inflation has to end before nucleosynthesis for successful structure formation. Otherwise, inflation would leave behind a cold universe empty of matter frozen in a state of low entropy. An era after inflation called reheating resolves this problem. During this stage the inflaton particles coupled to other particles can decay perturbatively and entropy production occurs \cite{Linde1}. As the perturbative approach does not involve the coherent nature of inflaton field, a non-perturbative particle production known as preheating phase was then proposed \cite{Kofman:1994rk,Kofman:1997yn,Felder:1998vq,Podolsky:2005bw,Bassett:2005xm,Allahverdi:2010xz}. In this phase particles are produced by parametric resonance mechanism in which the inflaton field begins to oscillate near the minimum of its effective potential and eventually produces many elementary particles. In this sense, in the second stage of inflaton decaying called reheating the produced particles interact with each other and come to a state of thermal equilibrium with some final temperature $T_{re}$ which usually corresponds to initial temperature of radiation epoch. Recently, it has been recognized that the reheating stage can constrain the inflationary models by calculating the reheating temperature as a function of spectral index $n_s$ \cite{Martin:2010,Creminelli:2014oaa,Dai:2014jja,Munoz:2014eqa,Creminelli:2014fca,Cook:2015vqa}.

In this work we introduce a new model for inflation, where a quadratic scalar potential (Eq.\eqref{1}) has been augmented with a non-minimal coupling $\xi$ to the Ricci scalar (Eq.\eqref{p1}) which satisfies the constraints of recent CMB data and reheating analysis. This model has another interesting properties such as a remnant energy density appropriate to describe the recent accelerated expansion of the universe which has recently received attention \cite{Liddle:2006qz,Liddle:2008bm,Henriques:2009hq,Lin:2009ta,Bose:2009kc,DeSantiago:2011qb,Ebrahimi:2012zz,Panotopoulos:2007ri,Cardenas:2007xh,Bastero-Gil:2015lga}. We start with a quadratic potential coupled non-minimally to the gravity in the Jordan frame and then move to the Einstein frame by employing a special conformal transformation. This transformation brings the gravitational and kinetic terms in the Lagrangian into canonical form and also gives the potential term a plateau-like shape. Then we apply the method recently developed in \cite{Martin:2010,Creminelli:2014oaa,Dai:2014jja,Munoz:2014eqa,Creminelli:2014fca,Cook:2015vqa} to put constraint on the non-minimal coupling $\xi$ of the model. In this approach, we calculate the reheating temperature $T_{\textrm{re}}$ and reheating e-folding number $N_{\textrm{re}}$ in terms of $n_s$, effective equation of state $\omega_{\textrm{re}}$ and $\xi$. Using the CMB constraint on $n_s$, the lower bound on the $T_{\textrm{re}}$ by primordial nucleosynthesis (BBN) and the upper bound on $T_{\textrm{re}}$ by the scale of inflation, it is possible to find a physical range for $\xi$. Indeed, the model makes distinct predictions for the parameters $n_s$ and $r$ as a function of $\xi$. At the next step, using the constraint placed on $\xi$ we compare the $n_s-r$ prediction of the model with the recent Planck results. We show that our model is consistent with the Planck data well.

 In the rest of the paper, we study the preheating mechanism of the model using a four-legs interaction between background inflaton scalar field and a quantum scalar field through non-linear preheating mechanism. We use and modify the LATTICEASY program for numerical simulations \cite{Felder:2000hq} to understand the microphysics in our system during perheating era. For a four-legs interaction between the noted scalar fields, the decay of the inflaton field is not complete. This mechanism which was first proposed in \cite{Kofman:1994rk}, involves the idea of incomplete inflaton decay that leaves part of the oscillating inflaton field decoupled, behaving as dark matter.

The paper is organized as follows: In section \rom{2}, we introduce the model and discuss its motivation. In section \rom{3}, we calculate the power spectrum of the model and check its consistency with Planck's new data. Also, we study the reheating constraints on the dynamics of its potential. In Section \rom{4} we study the preheating mechanism of the model using a four-legs interaction term. Finally, the conclusions of the present work are summarized in section \rom{5}.

\section{Building A Non-minimal Coupling Model in the Einstein Frame}

The Planck data \cite{Ade:2015lrj} shows that we live in a simple universe with small spatial curvature possessing a simple dynamical mechanism which caused the smoothing and flattening of the universe. These properties eliminate complicated inflationary models with multiple fields and non-canonical ones and in exchange justify single scalar fields.\\
\indent
On the other hand, we know observationally that the expansion of the universe is accelerating. The simplest explanation is supposing an extremely small but non-zero positive cosmological constant, $\Lambda$ which maybe introduced via the vacuum state of a scalar field with non-zero constant energy $V_0$ called dark energy. In string theory, there are a large number of vacua with zero or non-zero values $V(\phi)$ in which $\phi$ is the scalar field. Many of infinitesimal non-zero vacua which form a continuum lead to the creation of a large, observable universe via a self-reproducing inflationary mechanism. The space of all such vacua is called the landscape \cite{Susskind:2003kw}. There is a perfectly flat plain in the landscape called supermoduli-space where the vacua are super-symmetric and a quantum field potential $V(\phi)$ is zero. However, there are also possible non-zero values $V(\phi)$ which are stable or meta-stable local minima. Once, one moves off the supermoduli-space, the low energy properties of string theory which can be approximated by field theory may break down in some regions of the landscape and give a minimum energy for $V _0$ which may be interpreted as the cosmological constant. In this way, anthropic principle enforces that we live in a special part of the universe capable of galaxy formation and with infinitesimal non-zero $V_0$.\\
\indent
Therefore, combining a single scalar inflaton field potential with $V_0$ as
\begin{equation}
V(\phi)=V_0+\frac{1}{2}m_{\phi}^2 \phi^2~, \label{1}
\end{equation}
which is adopted in references like \cite{Liddle:2006qz,Cardenas:2007xh,Liddle:2008bm}, would be appropriate for explaining inflation and dark energy in a unified scenario. During inflation, the first term of \eqref{1} is negligible but at the late time that $\phi\ll 1$, it gives us accelerating expansion. In order for the fossil term $V_0$ to match the dark energy density, we have to set $V_0 = V_{\ast}\alpha=\rho_{DE}\simeq10^{-47}GeV^{4}$ \cite{Ade:2013zuv} where $\alpha$ is fine-tuned to be of order $\alpha\sim 10^{-112}$ and $V_{\ast}$ is the energy scale of inflation \cite{Ade:2015lrj} as
\beq
V_{\ast}\approx (1.88\times 10^{16}\:\textrm{GeV})^{4}\,\frac{r}{0.10}~.
\eeq
Cosmological evidence strongly suggests that the radiation and matter dominated eras of the cosmic evolution reside between two exponentially expanding phases: An early phase corresponding to cosmic inflation with a nearly constant Hubble parameter and a corresponding energy scale of the order of $10^{64} GeV^4$ and a second phase in which the universe has entered recently, as indicated by distant supernova observations. The recent quasi-inflationary period, however, is occurring at a much lower energy scale of the order of $10^{-47} GeV^4$. Besides the string theory motivation mentioned above, the main motivation for using the present potential is that it interpolates smoothly between these two hierarchical energy scales. It is interesting that such a potential originates from a surprisingly simple potential in the Jordan frame. However, there is still a problem with \eqref{1} as it includes a single scalar field potentials which is ruled out by recent Planck data \cite{Ade:2015lrj}. In what follows, we try to overcome this problem by the help of the non-minimal coupling scenario.\\
\indent
A general theory of inflation in which a single inflaton field couples non-minimally to gravity has been extensively studied during past years \cite{Futamase:1987ua,Fakir:1990eg,Kaiser:1994vs,Makino:1991sg,Komatsu:1997hv,Komatsu:1999mt,Bezrukov:2007ep,Kallosh:2013tua,Galante:2014ifa,Giudice:2014toa,Kallosh:2014rha}. In order to develop cosmological perturbations and calculate the power spectrum, it is usually useful to employ a conformal transformation technique to go from the Jordan frame to the Einstein frame in which the calculations are simpler. In the Einstein frame, we again get a minimally coupled theory with a new effective potential. Fortunately, it has been confirmed that the Jordan and Einstein frames describe the same physics and have equivalent predictions for the CMB anisotropies \cite{Fakir:1990eg,Futamase:1987ua,Kaiser:1994vs,Makino:1991sg,Komatsu:1997hv,Komatsu:1999mt,Hwang:1996np,Deruelle:2010ht,Postma:2014vaa,Kubota:2011re}. We can therefore move to the Einstein frame and apply the standard curvature perturbation computations. In general, one can consider the following action for the inflaton non-minimally coupled to gravity with non-canonical kinetic term \cite{Hwang:1995bv,Tsujikawa:1999iv,Tsujikawa:1999me,Starobinsky:2001xq,Noh:2001ia,Tsujikawa:2004my}
\bea
  S_J= \int d^4 x  \sqrt{-g} \left [ \frac{\Omega(\phi)}{2\,\kappa^{2}}\, R - \frac{\omega(\phi)}{2} \partial_\mu \phi
\,  \partial^\mu \phi  - V(\phi) \right]~,\label{action0}
\eea
where $R$ is the Ricci scalar, $V(\phi)$ is the potential term and $\Omega(\phi)$ and $\omega(\phi)$ are functions of the inflaton field. Moving to the Einstein frame by the following conformal transformation
\beq
\hat{g}_{\mu\nu}=\Omega(\phi)\,g_{\mu\nu}~,
\eeq
leaves the action \eqref{action0} in the following modified form
\beq
\label{action2} S_{E}= \int
d^4 x  \sqrt{-\hat{g}} \left [ \frac{1}{2\,\kappa^{2}}\, \hat{R} - \frac{1}{2}F^{2}(\phi)\hat{g}^{\mu\nu} \partial_\mu \phi
\,  \partial_\nu \phi   - \hat{V}(\phi) \right]~,
\eeq
with
\beq
F^2(\phi)=\frac{3}{2\kappa^{2}}\frac{\Omega'^{2}}{\Omega^{2}}+1~,
\eeq
and
\beq
\hat{V}(\phi)=\frac{V(\phi)}{\Omega^2},
\eeq
where we assumed $\omega(\phi)=\Omega(\phi)$. Later, we show that this choice is acceptable. Based on the universal attractor models, the generalized non-minimal coupling to gravity is described in Joran frame with $\Omega(\phi)=1+\xi f(\phi)$, where $f(\phi)$ can be an arbitrary function \cite{Kallosh:2013tua,Galante:2014ifa}. In this paper, our aim is to represent a single field plateau-like inflationary model describing inflation and dark energy which is consistent with recent CMB data. As explained before, the scalar potential in \eqref{1} which has a minimally coupling $\xi=0$ to gravity is ruled out by Planck recently. On the other hand, the $n_s - r$ plots in \cite{Ade:2015lrj} favor a special subclass of single field slow-roll inflationary models with plateau-like potentials and with a canonical kinetic term in the framework of Einstein gravity \cite{Ijjas:2013vea}. Therefore, to solve the problem of \eqref{1} and find a plateau-like potential, we use the non-minimally coupling scenario which helps us  by choosing an appropriate function $f(\phi)=(\xi)^{-1}(\sqrt{1+\kappa^2 \xi \phi^2}-1)$ which gives us
\begin{equation}
\Omega(\phi)^2 = 1+\kappa^2 \xi \phi^2 ~. \label{2}
\end{equation}
As we do not know the exact value of inflaton mass, we define $m_{\phi} = 2 V_{\ast} \kappa^{2}\xi$. In section \rom{3}, by using reheating analysis we put constraint on the value of $\xi$ and consequently on the value of $m_{\phi}$. Now by applying \eqref{1}, \eqref{2} and recent definition for $m_{\phi}$, we find $F$ and $\hat{V}$ as
\beq
F^2(\phi)=\frac{3}{2\kappa^{2}}\frac{(\kappa^{2}\xi\phi)^{2}}{(1+\kappa^{2}\xi\phi^{2})^{3}}+1~,
\eeq
and
\beq
\hat{V}(\phi)=V_{\ast}\frac{\alpha+\kappa^{2}\xi\phi^{2}}{1+\kappa^{2}\xi\phi^{2}}~. \label{p1}
\eeq

When inflation starts, we have $F\approx 1$ in the region $\phi^{2}\gg 1$ which gives us non-minimally coupled inflaton field in the Einstein frame. Therefore, the former choices for $\Omega(\phi)$ and $\omega(\phi)$ change the kinetic term of \eqref{action2} into a suitable canonical form.
At the end of inflation, the scalar field with the plateau-like potential \eqref{p1} has an oscillatory phase of evolution around $\phi =0$ and at the late time this potential is dominated by $\hat{V}(\phi)\approx V_{\ast}\alpha$ which can be interpreted as remnant dark energy \cite{Liddle:2006qz,Liddle:2008bm,Henriques:2009hq,Lin:2009ta,Bose:2009kc,DeSantiago:2011qb,Ebrahimi:2012zz,Panotopoulos:2007ri,Cardenas:2007xh,Bastero-Gil:2015lga}.\\
\indent
Someone may think about the similarity of our non-minimally coupled model, \eqref{p1} and  conformal $\alpha$-attractors \cite{Kallosh:2013hoa,Kallosh:2013daa,Ferrara:2013rsa,Kallosh:2013yoa,Kallosh:2014rga,Linde:2015uga}. The attractor nature of these theories comes from the existence of any leading pole in the kinetic term of their Lagrangian which lead to nearly the same predictions of cosmological observables in the limit of a large number of e-folds. Both classes have a free parameter and by choosing the parameter suitably one can make the model consistent with CMB data. However, these two classes differ except for a special case of the $\alpha$-attractors with a pole in the kinetic term \cite{Galante:2014ifa}.\\
\indent
In the next section we fix our free parameter $\xi$ using the constraint given by the reheating stage and then compare the cosmological predictions of \eqref{p1} with Planck data.

 \section{CMB constraints on the Inflationary fractional potential}

 In this section, we compare the predictions of the potential \eqref{p1} for $n_s$ and $r$ with the recent CMB data. At first we calculate the number of e-folding $N(\phi)$
 \begin{equation}
 N(\phi)=\kappa^{2}\int^{\phi }_{\phi_{e}}\frac{V}{V'}d\phi~.\label{N}
 \end{equation}
where $\phi_{e}$ is the value of inflaton field at the end of inflation. Integrating the expression \eqref{N} in the region long before the end of inflation, we arrive at
\beq
N(\phi)=\frac{\xi}{8}(\kappa\phi)^{4}~.
\eeq
Therefore, the slow roll parameters are also given in terms of $N$ as follows
\bea
\epsilon &\simeq &\frac{2}{\xi^{2}}\frac{1}{(\kappa\phi)^{6}}=\frac{1}{8\sqrt{2}\,\xi^{1/2}}\frac{1}{N^{3/2}}~,\\
 \eta &\simeq &-\frac{6}{\xi}\frac{1}{(\kappa\phi)^{4}}=-\frac{3}{4N}~.
\eea
Finally the $n_s$ and $r$ parameters are obtained as
\bea
n_s &\simeq & 1-\frac{3}{4\sqrt{2}\,\xi^{1/2}}\frac{1}{N^{3/2}}-\frac{3}{2N}~,\label{ns}\\
r &\simeq &\frac{\sqrt{2}}{\xi^{1/2}}\frac{1}{N^{3/2}}~,\label{r}
\eea
and hence we find the consistency relation
  \begin{equation}
 r=\frac{8}{3}\left(1-n_{s}-\frac{3}{2N}\right)~. \label{eqrer}
 \end{equation}

\begin{figure}
    \centering
        \subfigure[  ]
    {
        \includegraphics[width=3.4in]{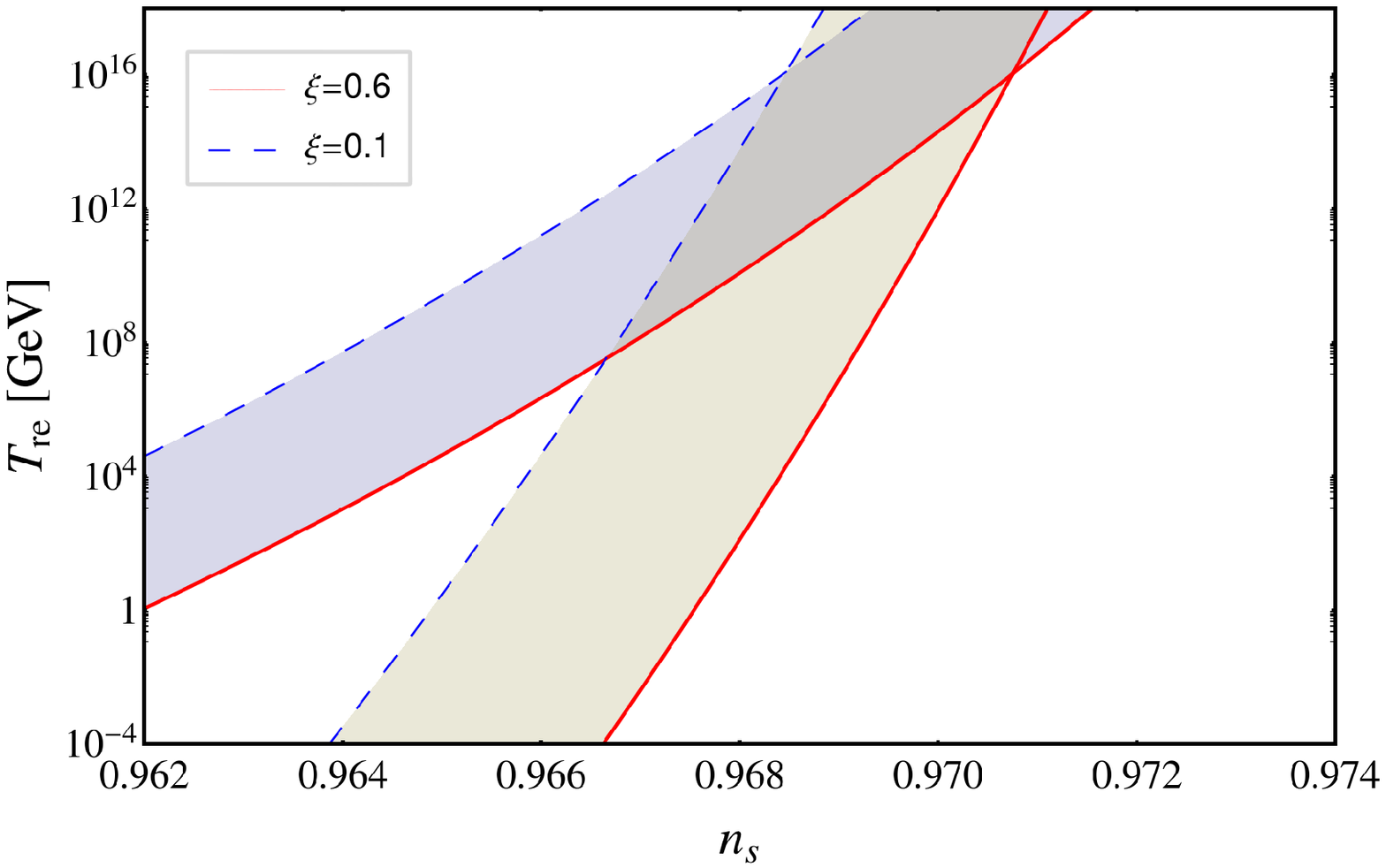}
        \label{fig:second_sub}
    }
       \subfigure[]
    {
        \includegraphics[width=3.4in]{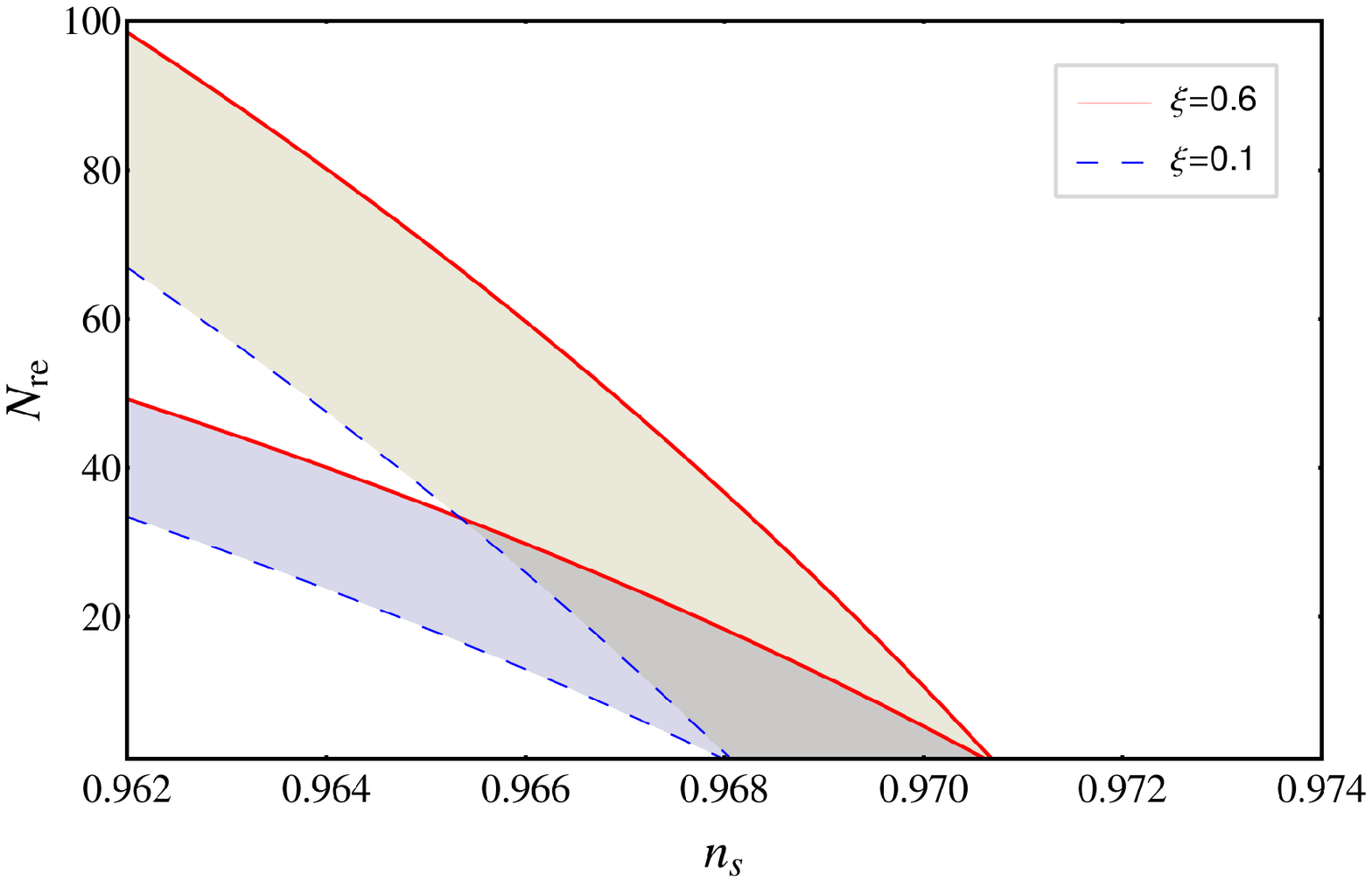}
        \label{fig:third_sub}
    }
    \caption{Plots of $T_{re}$ (a) and $N_{re}$ (b) as a function of $n_s$ in the 1$\sigma$ range from Planck \cite{Planck:2015xua}. In each plot, there are two allowable regions between blue dashed curve and red solid one which determine the upper and lower band of $T_{re}$ and $N_{re}$ for $\omega_{re}= 0$ (light blue region) and $\omega_{re}=1/6$ (gray region).}
    \label{fig:sample_subfigures}
\end{figure}

\begin{figure}\vspace{.5in}
      \includegraphics[width=4in]{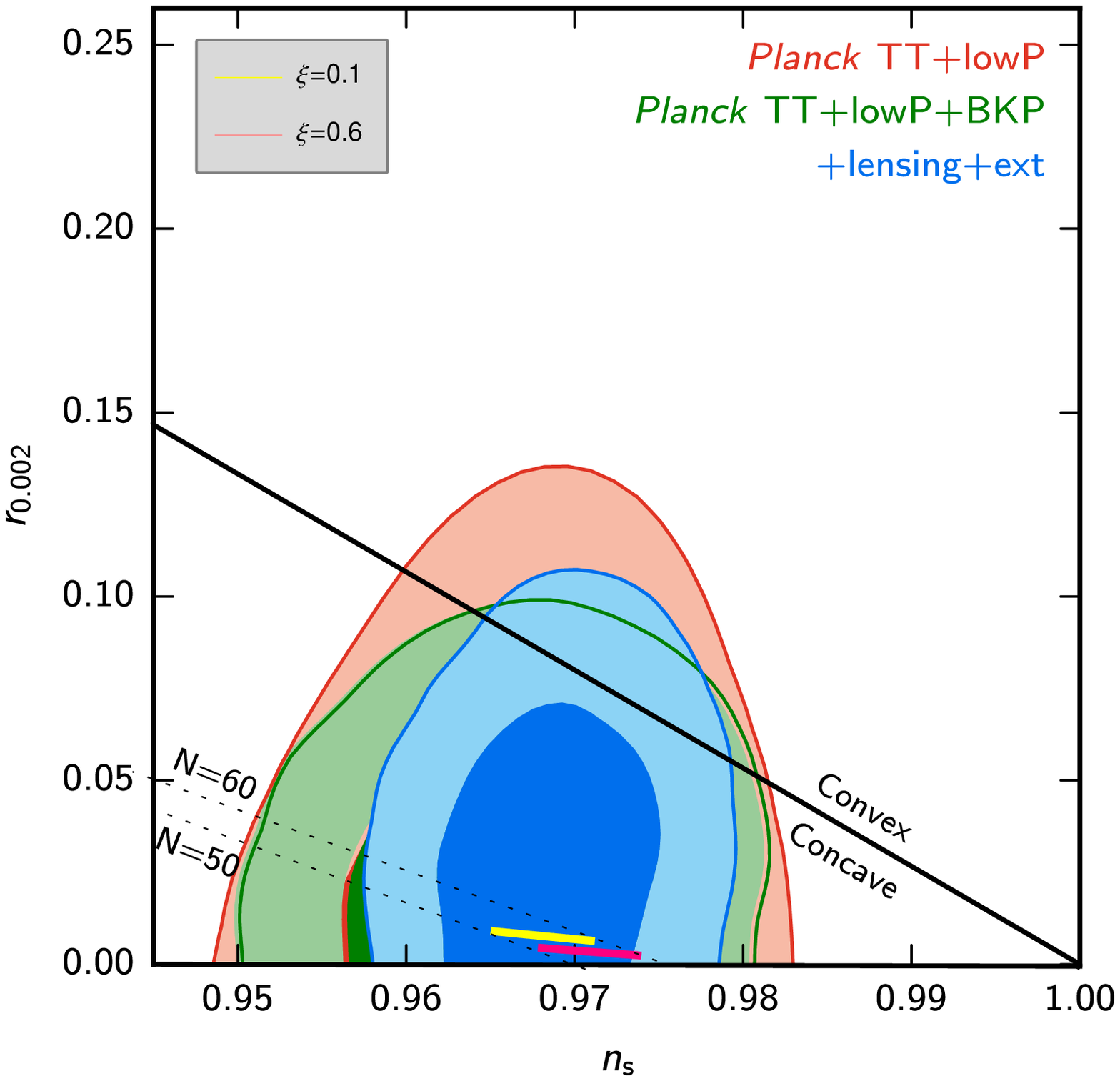}
    \caption{Constraints on $r$ for our fractional single field potential, using Planck TT+lowP (red region), BICEP2/Keck Array+Planck TT+lowP (BKP) (green region) and lensing+ext (blue region) which include B-mode polarization results \cite{Planck:2015xua}. Pink solid line ($\xi=0.6$) and yellow one ($\xi=0.1$) show lower and upper approximate $n_s-r$ relationship for this model assuming $50 < N < 60$. It is interesting that this model favors upper limit on $r_{0.002} < 0.11$ very well.}
    \label{fig2}
\end{figure}

Before comparing the model with data we discuss the possible constraints of reheating stage on the parameter $\xi$ using the methods proposed in \cite{Martin:2010,Creminelli:2014oaa,Dai:2014jja,Munoz:2014eqa,Creminelli:2014fca,Cook:2015vqa}. The reheating era begins once the inflation ends and the comoving horizon starts to increase. At this time the energy density of inflaton field is dissipated
 into a hot plasma with temperature $T_{\textrm{re}}$. The reheating can occur suddenly or it may take several e-folding times $N_{\textrm{re}}$. After this stage, the universe evolves under radiation domination regime for $N_{\textrm{RD}}$ e-folding numbers until the time of equality. Taking a mode with comoving momentum $k$, the amount of expansion during different eras are related as follow  \cite{Liddle:2003as}
  \begin{equation}
 \ln \frac{k}{a_0H_0}=-N_{k}-N_{\textrm{re}}-N_{\textrm{RD}}+\ln \frac{a_{\textrm{eq}}H_{\textrm{eq}}}{a_0H_0}+\ln \frac{H_{k}}{H_{\textrm{eq}}}~ ,\label{eqre1}
 \end{equation}
where the subscripts refer to different eras, including the horizon exit ($k$), reheating (re), radiation (RD), radiation-matter equality (eq) and the present
time (0). During inflation we have $H^{2}_{k}=\pi^{2} rA_s/(2\kappa^{2})$ with the primordial scalar amplitude $\ln (10^{10} A_s)=3.089$ \cite{Ade:2013zuv}.
 Using the continuity equation one can write
 \begin{equation}
 N_{\textrm{re}}=\frac{1}{3(1+\omega_{\textrm{re}})}\ln \left(\frac{\rho_{\textrm{end}}}{\rho_{\textrm{re}}}\right)~,\label{eqre2}
 \end{equation}
where $\omega_{\textrm{re}}$ is the reheating equation of state parameter and $\rho_{\textrm{end}}=(1+\delta)V_{\textrm{end}}$ is the inflaton energy density at the end of inflation with $\delta$ the ratio of the kinetic energy to the potential energy. The results of the following calculations and their related plots are not sensetive to value of $\delta$. Also we have $\rho_{\textrm{re}}=(\pi^2/30)g_{\textrm{re}}T_{\textrm{re}}^4$  with $g_{\textrm{re}}$ being the effective degrees of freedom of relativistic species
upon thermalization. On the other hand, one can relate the reheating temperature $T_{\textrm{re}}$ to the present CMB temperature $T_0$ through
 \begin{equation}
 \frac{T_{\textrm{re}}}{T_0}=\left(\frac{43}{11g_{\textrm{s},\textrm{re}}}\right)^{1/3}\frac{a_0}{a_{\textrm{eq}}}\frac{a_{\textrm{eq}}}{a_{\textrm{re}}}~ ,\label{treh1}
 \end{equation}
where $g_{\textrm{s},\textrm{re}}$ is the effective number of light species for entropy at reheating. Combining equations \eqref{eqre1} and \eqref{eqre2} and using \eqref{eqrer}, we obtain the e-folding number of reheating for our model as
 \begin{eqnarray}
 N_{\textrm{re}}=\frac{4}{1-3\omega_{\textrm{re}}}\displaystyle \left[- \ln \left(\frac{k}{a_0 T_0}\right)-N_k-\frac{1}{4}\ln \left(\frac{30}{\pi^2 g_{re}}\right)
- \frac{1}{3}\ln\left(\frac{11 g_{s,re}}{43}\right) \right. \nonumber\\
 \left. - \frac{1}{4}\ln(1+\delta)
+\frac{1}{4}\ln \left(\frac{\pi^2rA_s}{6}\right)-\frac{1}{4}\ln \left(\frac{(2\xi)^{1/3}}{1+(2\xi)^{1/3}}\right) \right ]~,\label{Nreh}
 \end{eqnarray}
Moreover, putting the definitions of $\rho_{\textrm{re}}$ and $\rho_{\textrm{end}}$ into \eqref{eqre2} one can calculate the reheating temperature in the form
 \begin{equation}
 T_{re}=\frac{1}{\kappa}\left[\frac{45(1+\delta)rA_s}{g_{re}}\frac{(2\xi)^{1/3}}{1+(2\xi)^{1/3}}\right]^{1/4} \exp \left(-\frac{3}{4}(1+\omega_{re})N_{re}\right)~.\label{Treh}
 \end{equation}
We set the values $g_{re}=g_{s,re}=100$ and plot (\ref{Treh}) and (\ref{Nreh}) in Fig. 1 (a) and (b) in terms of $n_s$ with the range within the $1\sigma$ region allowed by the Planck results around $n_{s}=0.968\pm 0.006$ \cite{Planck:2015xua}. From upper and lower limits on $T_{re}$ we find that a value of $\omega_{re}$ out of the range $0\leq\omega_{re}\leq1/6$ is disfavored from model building.
Therefore, in these plots the results are depicted only for $\omega_{re}=0$ and for $\omega_{re}=1/6$. If we fix the upper bound on $T_{re}$ to be the scale of GUT $\sim 10^{16}\:\textrm{GeV}$ and the lower bound on $T_{re}$ to be the scale of BBN $\sim 10^{-2}\:\textrm{GeV}$ \cite{Steigman:2007xt} then the physical range for the non-minimal coupling parameter $\xi$ is restricted by $0.1\lesssim \xi \lesssim0.6$. Hence, the width of each curve in either Fig. 1 (a) and (b) corresponds to this bound on the parameter $\xi$. Our reheating results of the fractional potential model for the central value $n_s\simeq 0.968$ shows that in the case of $\omega_{re}=0$, a number $N_{re}\lesssim20$ of e-folds and $T_{re}\gtrsim10^{10}\:\textrm{GeV}$ is required, while for $\omega_{re}=1/6$ a number $N_{re}\lesssim37$ of e-folds and $T_{re}\gtrsim10^{2}\:\textrm{GeV}$ is favored.

In the next step, using the bound obtained on the parameter $\xi$ we can compare the predictions of the fractional potential model for the $n_s(\xi)$ and $r(\xi)$ calculated in \eqref{ns} and \eqref{r} with the recent Planck data. Fig. 2 shows that the prediction of this potential is in very good agreement with the $n_s - r$ plane favored by the Planck data \cite{Planck:2015xua,Ade:2015lrj}. This agreement is achieved for the bounds $\xi=0.1$ and $\xi=0.6$ which were consistent with the reheating constraints. As one can see, the model predicts $0.004\leq r\leq 0.012$ for the gravitational waves which may be detected in future experiments.

\section{Preheating Phase}

As it was pointed out in section \rom{1}, at the beginning of the reheating era the inflaton field begins to oscillate near the minimum of its effective potential and consequently approaches a sinusoidal solution with decreasing amplitude. Reheating was initially studied by employing the usual perturbative approach \cite{Dolgov:1982mp,Abbott:1982hn}. However, it was shown by \cite{Shtanov:1994ce,Kofman:1994rk,Kofman:1997yn} that the perturbative analysis does not take into account the coherent nature of the inflaton field. In general the decay of inflaton field can be started earlier in a preheating stage where inflaton decays non-perturbatively. In this phase, particles are produced through the parametric resonance mechanism \cite{Kofman:1994rk,Kofman:1997yn,Felder:1998vq,Podolsky:2005bw,Bassett:2005xm,Allahverdi:2010xz}. In this way, the coherent effects of the scalar field can lead to the decaying of the homogenous field much faster than the perturbative method. In order to study the preheating mechanism we consider a four-legs interaction of the form
\begin{equation}
\mathcal{L}_{\textrm{int}}=-\frac{1}{2}g^2\phi^2\chi^2~,
\end{equation}
where $g $ is a dimensionless coupling constant and $\chi$ is a quantum scalar field coupled to the classical inflaton field. In the external inflaton background one can decompose the quantum field $\chi$ into
\begin{equation}
\chi(t,\mathbf{x})=\frac{1}{(2\pi)^{3/2}}\int d^{3}k
\left(\chi_{k}^{*}(t)\hat{a_{k}}e^{i\mathbf{k}\cdot\mathbf{x}}+\chi_{k}(t)\hat{a}_{k}^{\dag}e^{-i\mathbf{k}\cdot\mathbf{x}}\right)~,
\end{equation}
where $\hat{a_{k}}$ and $\hat{a}_{k}^{\dag}$ are the creation and annihilation operators, respectively.
The Klein-Gordon equations of motion for the coupled fields $\phi$ and $\chi$ in an expanding FRW universe are
\begin{eqnarray}
&&\ddot{\phi}+3H\dot{\phi}-\frac{1}{a^2}\nabla^2 \phi+ V'+g^2\chi_{k}^2\phi=0~,\label{eqre3}\\
&&\ddot{\chi}+3H\dot{\chi}-\frac{1}{a^2}\nabla^2 \chi+g^2\phi^2\chi=0~,\label{eqre4}
\end{eqnarray}
and the evolution of scale factor is given by Friedmann equations. The equations of motion of the both fields are oscillatory equations with effective comoving frequency $\omega_{k}^2 = k^2+m_{\textrm{eff}}^2\, a^2$. Taking $m_{\chi}\ll m_{\phi}$ during inflation, the effective masses of the fields will be $m_{\phi,\,\textrm{eff}}^2= m_{\phi}^2+g^2\langle\chi^2\rangle$ and $m_{\chi,\,\textrm{eff}}^2= g^2\langle\phi^2\rangle$, respectively. Assuming an inflaton field with mass $m_{\phi}$ oscillating with the amplitude $\Phi(t)$, one can usually define the dimensionless parameters $q = (g^2 \Phi(t)^2)/(4m_\phi^{2})$ and $A_k=2\,q+k^2/(m^2a^2)$ characterizing the parametric amplification of $\chi_{k}$ modes. For instance in an expanding universe with $q\gg 1$ one can find a phase of broad parametric resonance where at each oscillation of the field $\phi$, the field $\chi$ oscillates several times with sharp peaks in the density number of particles $n_k$ \cite{Kofman:1994rk}. The limits on the coupling $g^2$ which have been discussed in \cite{Podolsky:2005bw} lead to $10^{-3} \lesssim q \lesssim 10^5$. Indeed, we should find an optimal value for $g$ large enough to produce highly efficient preheating, but small enough that the occupation numbers $n_k \sim 1/g^2$ produce strong re-scattering.
For $g\phi_{0}<m_{\varphi}$ there is a resonance with $q\ll 1$ in a narrow instability band about $k=m$ called narrow parametric resonance. In this regime of resonance, for each oscillation of the field $\phi$ the growing modes of the field $\chi_{k}$ oscillate one time. For $q \gg 1$, i.e. for oscillations with a large amplitude $\phi_{0}$, the resonance occurs for a broad range of $k$. Therefore, the broad parametric resonance is extremely more efficient \cite{Kofman:1994rk}.

In this section we turn to study the preheating mechanism of the potential \eqref{p1} through the broad parametric resonance regime. In our case when $\phi\rightarrow0$ and hence inflaton oscillates around its minimum, we can expand \eqref{p1} in the form
\beq
\hat{V}(\phi)\simeq V_{\ast}\left(\alpha+\kappa^2\xi\phi^2-\kappa^4\xi^2\phi^4+\cdot\cdot\cdot\right)~,\label{pot1}
\eeq
 where $\alpha$ is infinitesimal and $\xi \phi \ll1$ during reheating era. Hence, we keep only the second and third terms for our preheating calculations. This tree-level potential \cite{Sahni:2004ai} has a stable minimum at $\phi=0$ and the system eventually settles down in this ground state during time evolution. Now, by using the potential \eqref{pot1} we can study the dynamics of the fields $\phi$ and also $\chi$ after inflation through non-linear preheating mechanism. Here we modify LATTICEASY program \cite{Felder:2000hq} which solves \eqref{eqre3} and \eqref{eqre4} and also Friedmann equation on a three-dimensional lattice using finite differencing for spatial derivatives and a second order staggered leapfrog algorithm for time evolution. In our choice, the comoving edge size of the cubic lattice is $L=20/m$, number of points along each edge is $N=256$, and the size of time step of calculation is $m\,dt=0.001$. For the rest of this paper all simulations are performed with $\phi_0 = 0.139 M_P$, $m_{\phi}=10^{-6}M_P$ and $g^2=2.5\times 10^{-6}$ which is the optimal value. The $\xi$ was constrained using the reheating constraint in the previous section. Therefore, we set $\xi=0.1$ for our simulation.

 At the first step, we consider the total energy densities, $\varepsilon$ of the both fields $\phi$ and $\chi$ containing kinetic and the gradient energies plus their interaction term. Fig. 3 compares the contributions of these energy components of our potential \eqref{pot1} with equivalent terms of the quadratic potential $m^2\phi^{2}/2$. For the chaotic potential we have considered $m=10^{-6}\,M_{Pl}$. In this plot the black curves correspond to the potential \eqref{pot1} and the red ones are for the quadratic potential. All of the blue curves show the energy density of interaction term, $g^2\phi^2\chi^2/2$, in the potential \eqref{pot1}. As one can see, two curves exhibit slightly different behavior. It is significant that the contribution from inflaton field is dominant even after preheating, up to approximately $m\,t\sim 175$. We find that the interaction term (blue curve), except for a temporary time interval $80 \leqslant m\,t \leqslant 150$, is sufficiently small that allows us to define energy density for each field in the following form
\beq
\varepsilon \approx \frac{1}{(2\pi)^3 a^4} \int d^3k\, \omega_k \,n_k~,
\eeq
where $\omega_k$ is the co-moving frequency and $n_k$ is the well defined co-moving occupation number
\beq
n_k= \frac{\omega_k}{2}\left(\frac{|\dot{X}_k|^2}{\omega_k^2}+|X_k|^2\right)-\frac{1}{2}~,
\eeq
with $X_k(t)=a^{-3/2}\chi_k$ \cite{Podolsky:2005bw}. There are three stages for the dynamics that govern the preheating phase: linear dynamics, nonlinear dynamics and perturbative dynamics. At linear dynamical regime, through early re-scattering the $\chi$ particles are created in the resonant band and then generation of $\delta\phi$ quanta occurs by the annihilation process $\delta\chi_k\delta\chi_k\rightarrow\delta\phi_k\delta\phi_k$. The second stage which its evolution is under a nonlinear dynamics, starts at $mt\sim100$ and ends at $mt\sim150$. In this stage, the particle occupation number is not well-defined because the amplitude of interaction potential is comparable to other contributions of total energy density. At the third stage, the energy density contributions evolve smoothly toward the higher comoving momenta under a perturbative dynamics.

\begin{figure}\vspace{.5in}
      \includegraphics[width=4in]{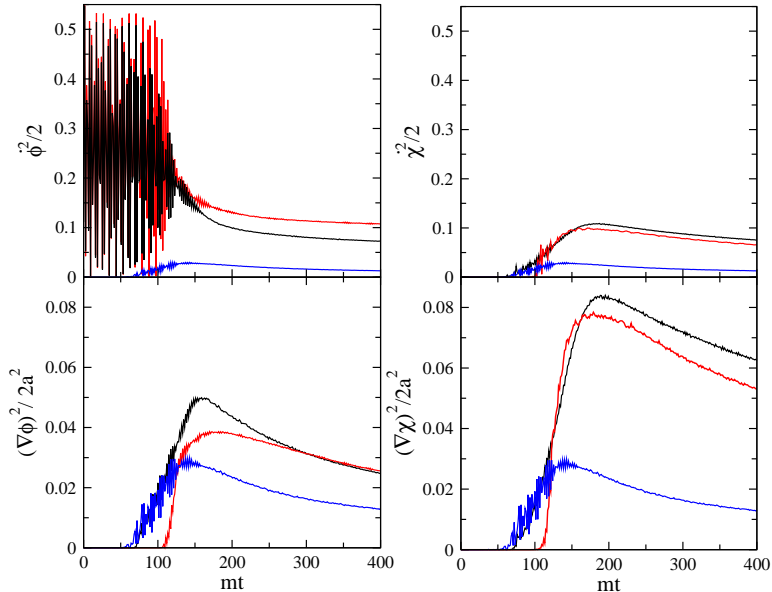}
    \caption{Contributions of kinetic and gradient energies of fields $\phi$ (left plots) and $\chi_{k}$ (right plots) in comparison with quadratic potential. The black curves belong to our model while the red ones relate to quadratic model. Also the blue one in each plot represent the energy density of interaction term of our potential.}
   \label{fig3}
\end{figure}

Now we turn to the time evolution of occupation number spectra $n^\phi _k$ and $n^\chi _k$ as functions of $k/m$. We compare the time evolution of these spectra of the potential \eqref{pot1} with the ones of the quadratic potential in Fig. 4. In this figure the black curves show the $n^{\phi,\chi} _k$ of our potential and the red ones show the $n^{\phi,\chi} _k$ computed for the quadratic potential. The single green curves in each plot also show the $n^{\phi,\chi} _k$ spectra of the potential \eqref{pot1} at the final time $m\,t_{f}=200$. The spectra show the exponential growth with a resonant peak at $k=k_{\ast}$ through parametric resonance. As we can see, the amplitude of the spectra for the potential \eqref{pot1} has decreased with respect to the quadratic potential. The spectra move down for the modes $k<k_{\ast}$ due to the re-scattering process. The green curves show that amplitude of $n^\phi _k$ decreases faster than $n^\chi _k$ at the end of preheating. People usually believe that preheating is followed by a reheating era leading to complete decay of the inflaton field. However, for a four-legs interaction, $g^2\phi^2\chi^2$ we applied in this paper, the inflaton field does not decay completely. We know that if one considers a three-legs coupling between the inflaton field and other quantum fields \cite{Kofman:1994rk,Kofman:1997yn,Podolsky:2005bw}, the inflaton decays completely. It is also possible to consider the incomplete decay as a mechanism for generating dark matter \cite{Cardenas:2007xh,Panotopoulos:2007ri}.

\begin{figure}\vspace{.5in}
      \includegraphics[width=4in]{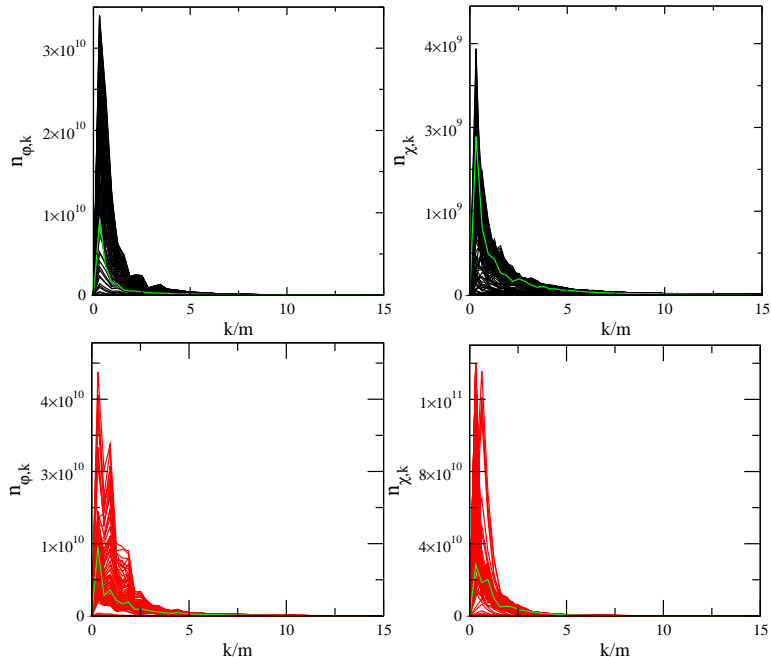}
    \caption{Time evolution of $n^\phi _k$ (left plot) and $n^\chi _k$ (right plot). The black curves are related to our potential while the red ones are for quadratic model. The green curves are the spectrum at the final time of our simulations.}
    \label{fig4}
\end{figure}

\section{Conclusion }

In this work, we introduced a new model for inflation by considering a reasonable non-minimal coupling model in Jordan frame and then moving to the Einstein frame through conformal transformation. The arising potential in Einstein frame had a plateau-like shape with a free non-minimal coupling parameter $\xi$. We studied this model extensively. First, we fixed the allowed range $0.1\leq\xi\leq 0.6$ using the constraints on the reheating temperature for a favored range of $0\leq\omega_{re}\leq1/6$ in Fig. \ref{fig:sample_subfigures}. Then in Fig. \ref{fig2}, using the constrained $\xi$ we compared $n_s(\xi)$ and $r(\xi)$ parameters predicted by this model with the recent CMB data released by Planck collaboration \cite{Planck:2015xua}. We found that this model predicts small tensor-to-scalar ratio of the order of $r\approx 0.01$ which is favored by the recent data very well. In the next step, we studied the out-of-equilibrium nonlinear dynamics of the interacting fields during preheating by the use of lattice numerical simulations. In this era, oscillations of $\phi$ lead to parametrically resonant amplification of $\chi_k$ and in turn the amplified modes of $\chi_k$ excite fluctuations of $\phi$. At the next stage of preheating that the interaction term becomes sub-dominant, we studied the evolution of the components of the total energy density and also the total occupation number as functions of time and compared them with equivalent terms of the quadratic model of inflation in Figs. \ref{fig3} and \ref{fig4}. One can see that $n^\phi _k$ grows exponentially faster than $n^\chi _k$ at the beginning of preheating due to rescattering and conversely its amplitude decreases faster than $n^\chi _k$ at the end of preheating.

\section*{\small Acknowledgement}

The authors would like to thank A. Linde, G. Felder, P. Creminelli, A. Abolhasani and Sh. Dehdashti for various comments and useful discussions. Mehdi Eshaghi acknowledges C. Baccigalupi for his helpful comments and thanks the organizers of Astrophysics sector of SISSA for their hospitality during the completion of this work.


\end{document}